%% file: arxiv_hip21152_letter_v2.tex
\documentclass[twocolumn]{aastex63}
\usepackage{graphicx}
\usepackage{enumerate}
\usepackage{amssymb, amsmath}
\usepackage{natbib}
\usepackage{color}
\usepackage{ulem}
\usepackage{dblfnote}
\usepackage{appendix}
\usepackage{hyperref}
\usepackage[stable]{footmisc}
\usepackage{relsize}
\usepackage{lineno}

\definecolor{red}{rgb}{1.0,0.0,0.0}

\begin{document}

\title{Direct Imaging Discovery and Dynamical Mass of a Substellar Companion Orbiting an Accelerating Hyades Sun-like Star with SCExAO/CHARIS}

\input{authors}

\shorttitle{Direct Imaging of HIP 21152 B}
\shortauthors{Kuzuhara et al.}

\begin{abstract}
We present the direct-imaging discovery of a substellar companion in orbit around a Sun-like star member of the Hyades open cluster. So far, no other substellar companions have been unambiguously confirmed via direct imaging around main-sequence stars in Hyades. The star HIP 21152 is an accelerating star as identified by the astrometry from the Gaia and Hipparcos satellites. We have detected the companion, HIP 21152 B, in multi-epoch using the high-contrast imaging from SCExAO/CHARIS and Keck/NIRC2. We have also obtained the stellar radial-velocity data from the Okayama 188cm telescope. The CHARIS spectroscopy reveals that HIP 21152 B's spectrum is consistent with the L/T transition, best fit by an early T dwarf. Our orbit modeling determines the semi-major axis and the dynamical mass of HIP 21152 B to be 17.5$^{+7.2}_{-3.8}$ au and 27.8$^{+8.4}_{-5.4}$ $M_{\rm{Jup}}$, respectively. The mass ratio of HIP 21152 B relative to its host is $\approx$2\%, near the planet/brown dwarf boundary suggested from recent surveys. Mass estimates inferred from luminosity evolution models are slightly higher (33--42 $M_{\rm{Jup}}$). With a dynamical mass and a well-constrained age due to the system's Hyades membership, HIP 21152 B will become a critical benchmark in understanding the formation, evolution, and atmosphere of a substellar object as a function of mass and age. Our discovery is yet another key proof-of-concept for using precision astrometry to select direct imaging targets.

\end{abstract}

\keywords{Brown dwarfs (185); Exoplanets (498); Open star clusters (1160); Direct imaging (387); Astrometry (80); Exoplanet detection methods (489); Astronomical instrumentation (799); Coronagraphic imaging (313)}

\section{Introduction}
The direct imaging (DI) technique is capable of detecting substellar companions with masses comparable to $\sim$1--20 Jupiter masses ($M_{\rm{Jup}}$) at projected separations wider than approximately 10 au, as demonstrated by discoveries such as the planets around HR 8799, $\beta$ Pic, 51 Eri, PDS 70, and AB Aur \citep[e.g.,][]{Marois2008a, Lagrange2010,Macintosh2015,Keppler2018, Currie_2022_ABAur}.
However, extensive volume/age limited DI surveys have revealed a low ($<$ 10 \%) occurrence rate for planet-mass companions \citep[e.g.,][]{Nielsen2019}.

Recent work shows the advantage of targeting stars that show evidence for the dynamical pull of a substellar companion, which provides a complementary approach to \textit{blind} surveys.
For example, targeted high-contrast imaging observations of the nearby Sun-like star HD 33632A from the \textit{Hipparcos}-\textit{Gaia} Catalogue of Accelerations \citep[HGCA;][]{Brandt2018,Brandt2021_HGCA} have revealed a brown dwarf (BD) companion in the system \citep{Currie2020}.
The HGCA lists all nearby stars with significant proper motion (PM) accelerations and allows to select promising targets for high-contrast imaging, since the accelerated PM of a star can be caused by its companion.
In addition, the HGCA is useful for analyzing the orbits of companions by combining it with DI data and/or radial velocity (RV) measurements, often leading to a $\sim$10\% dynamical constraint on the companion's mass \citep[e.g.,][]{Currie2020,Bowler_2021_HD47127,Brandt_2021_ORVARA}.
Thus, the use of HGCA also enables placing constraints on stellar and substellar evolution models by comparing the model-based mass of a companion with its dynamical mass measurement. \par

Imaged substellar companions around accelerating stars become even better benchmark objects if key system properties such as age and metallicity are well determined.
The Hyades open cluster is one of the most extensively examined open clusters (OCs) in all of astronomy, with a thoroughly vetted membership list, well constrained age, and well determined metallicity \citep[e.g.,][]{Brandt2015,Gagne2018,Gossaage_2018_Hyades}.
With typical distances of about 50 pc, Hyades members are near and bright enough that HGCA is well suited for identifying substellar companions.

We report the discovery of an L/T-transition BD companion around the accelerating star HIP 21152\footnote{The discovery of HIP 21152 B was independently reported by \citet{Bonavita_2022_COPAINS} with their VLT/SPHERE imaging performed among a survey for a large sample of accelerating targets. Franson et al. (in prep) have also independently discovered HIP 21152 B with their originally obtained data, and will characterize this system in detail with all the available data.}, with the companion’s dynamical mass estimation.
It is the first substellar companion directly imaged around a Sun-like star in the well-characterized Hyades OC and represents a new benchmark to better understand the properties of substellar objects.

\begin{deluxetable*}{llllllllll}
     \tablewidth{0pt}
    \tablecaption{HIP 21152 Observing Log and Companion Positions\label{obslog_hip21152}}
    \tablehead{
    \colhead{UT Date (MJD\tablenotemark{$\dag$})} &
    \colhead{Instrument\tablenotemark{$\ddagger$}} &
    \colhead{$\theta_{\rm{v}}$} &
    \colhead{$t_{\rm exp}$} &
    \colhead{$N_{\rm exp}$} & \colhead{$\Delta Par$} &
    \colhead{Data} &
    \colhead{S/N} &
    \colhead{$\rho$} &
    \colhead{PA}
    \\
    \colhead{} &
    \colhead{} &
    \colhead{(\arcsec{})} &
    \colhead{(s)} &
    \colhead{} &
    \colhead{($\arcdeg$)} &
     \colhead{Proc.} &
     \colhead{} &
     \colhead{(mas)} &
     \colhead{(\arcdeg)}
    }
    \startdata
    2020-10-07 (59129.589) & SCExAO/CHARIS & 0.4--0.6 & 25.08 & 235 & 85.6 & ADI & 19.3 & 408.5 $\pm$ 4.5 & 217.40 $\pm$ 0.66 \\
    2020-12-04 (59187.445) & SCExAO/CHARIS & 0.5--0.7 & 25.08 & 131 & 62.6 & ADI & 15.8 & 401.4 $\pm$ 4.5 & 216.66 $\pm$ 0.69 \\
    2020-12-25 (59208.380) & PyWFS+NIRC2 &  0.5--0.6 & 60 & 65 & 60.7 & ADI & 10.7 & 406.2 $\pm$ 6.0 & 216.39 $\pm$ 0.85 \\
    2021-10-14 (59501.591) & SCExAO/CHARIS & 0.6--0.7 & 30.98 & 41 & 17.4 & ASDI & 10.0 & 378.7 $\pm$ 5.1 & 216.90 $\pm$ 0.79 \\
    \enddata
    \tablenotetext{\dag}{Center epochs (modified julian days) during total exposure sequences.}
    \tablenotetext{\ddagger}{The wavelength range for CHARIS is 1.16--2.37 $\mu$m, while the $L^{\rm{\prime}}$-filter's central wavelength for NIRC2 is 3.78 $\mu$m.}
    \tablecomments{
    $\theta_{\rm{v}}$ represents the characteristic seeing measurements from the Canada France Hawaii Telescope seeing monitor.
    The integration time of each exposure, the numbers of exposures used in our analysis, the total variation of parallactic angle in each sequence are represented by $t_{\rm exp}$, $N_{\rm exp}$, and $\Delta Par$, respectively.
    The column of ``Data Proc.'' describes the types of our ALOCI PSF subtractions.
    S/N represents the companion PSFs' signal-to-noise ratios calculated from the 22-channel collapsed images.}
\end{deluxetable*}

\section{HIP 21152 System Properties, Observations, and Data}\label{sec: target_obs}

HIP 21152 (HD 28736) is a nearby \citep[$d$ = 43.208$^{+0.050}_{-0.049}$ pc;][]{Bailer-Jones_2021_Distance} F5V star \citep{Hoffleit1964} with an estimated mass of $\sim$ 1.3 $M_{\odot}$ \citep{David2015}.
For this star, we first found a substantial deviation from simple linear kinematic motion (i.e., acceleration) from the HGCA based on Gaia DR2 \citep{Brandt2018}.
The updated measurement of acceleration in the HGCA based on Gaia EDR3 \citep{Brandt2021_HGCA} is calculated to be $\chi^{2}$ = 174.6, consistent with a 13.0-$\sigma$ significance with 2 degrees of freedom (2 DOF).
The Banyan-$\Sigma$ \citep{Gagne2018} algorithm\footnote{\url{http://www.exoplanetes.umontreal.ca/banyan/}} provides HIP 21152 an extremely high membership probability (99.5\%) for Hyades with the inputs from the Gaia EDR3 catalogue \citep{Gaia_2021_Summary}.
The age of the Hyades OC was calculated to be 750 $\pm$ 100 Myr by \cite{Brandt2015} and 676$^{+67}_{-11}$ Myr\footnote{This is one of the six results in \cite{Gossaage_2018_Hyades}.} by \cite{Gossaage_2018_Hyades}, taking stellar rotations into account.

\subsection{SCExAO/CHARIS and Keck/NIRC2 High-Contrast Imaging} \label{sec: obs_reduction}

We performed high-contrast imaging observations using adaptive optics (AO) on the Subaru and Keck II telescopes between October 2020 and October 2021, in photometric and good-to-average seeing nights (Table \ref{obslog_hip21152}).
The Subaru observations utilized AO188 \citep{Hayano_2008_AO188} for first stage correction of atmospheric turbulence, followed by a faster higher order correction of residual wavefront errors by the extreme AO system SCExAO \citep{Jovanovic2015,Currie2020b}.
A coronagraph within SCExAO is then deployed to mask the central starlight, yielding high contrast images that are captured by the CHARIS integral field spectrograph \citep[IFS;][]{Groff2016}.
In our Keck observations, the target lights corrected by a near-IR Pyramid wavefront sensor were transferred to the NIRC2 camera \citep{Bond2020}.

The SCExAO/CHARIS and Keck/NIRC2 data obtained in 2020 have 55--98 minutes of on-source integration time.
Our shallower SCExAO/CHARIS data set obtained in October 2021 aimed solely at rejecting the possibility that the companion candidate is a background object.
All observations were performed in \textit{angular differential imaging} (ADI) mode, and CHARIS's IFS also enabled spectral differential imaging \citep[SDI; see][and references therein]{Oppenheimer_Hinkley_review_2009}.
All CHARIS data were taken with a low-resolution ($\mathcal{R}$ $\sim$ 18) spectroscopic mode to obtain wide wavelength coverage and the Lyot coronagraph with a 0\farcs{}23 diameter mask; NIRC2 data were taken in the $L^{\rm \prime}$-band filter also using a Lyot coronagraph but with a larger (0\farcs{}6) mask (see Table \ref{obslog_hip21152}).
By modulating SCExAO's deformable mirror, we generated four satellite spots around the point spread function (PSF) of HIP 21152 to enable astrometric and spectrophotometric calibration \citep[][]{Sahoo_2020_satellite}, while the NIRC2 coronagraph allows a direct stellar centroid estimate due to its partial transparency.
To flux-calibrate the NIRC2 data, we obtained unsaturated PSFs of the star before and after the coronagraphic sequence.

Using the pipeline of \citet{Brandt2017}, the raw CHARIS data were calibrated, and converted into 2D-image cubes consisting of 22 wavelength channels.
To further process these extracted data cubes, we used the CHARIS Data Processing Pipeline\footnote{\url{https://github.com/thaynecurrie/charis-dpp}} following the outline in \citet{Currie2020b}; the details of our high-contrast image processing are provided in Appendix \ref{app: detail_highcontrast}.
The spectra of HIP 21152A were measured with the satellite spots in each channel for spectrophotometric calibration, where we adopted an F5V model atmosphere from the Kurucz library \citep{Castelli2003} with the star's 2MASS photometry \citep{Skrutskie_2006_2MASS,doi-2MASS}.
The satellite spots are also used to register the central star's PSFs to a common center.
While the four spots have roughly equal brightnesses in each wavelength slice, the spot intensities for the December 2020 data showed a large systematic variation, making spectrophotometric calibration for these data more uncertain (see Appendix \ref{app: spec_dec}).
We processed the NIRC2 images using a well-tested pipeline \citep{Currie2011,Currie2014_HR8799} that carries out standard steps of sky subtraction, image registration, photometric calibration, and PSF subtraction.
Our image processing primarily adopts the ALOCI algorithm for ADI PSF subtractions \citep[e.g.,][]{Currie2014_HR8799,Currie2018}.
We attempted additional processing to improve or validate the fiducial reductions using alternate ADI reductions that adopt proprietary version of ALOCI and SDI speckle suppression (see also Appendix \ref{app: detail_highcontrast}).

\begin{figure*}
\includegraphics[width=1.0\textwidth]{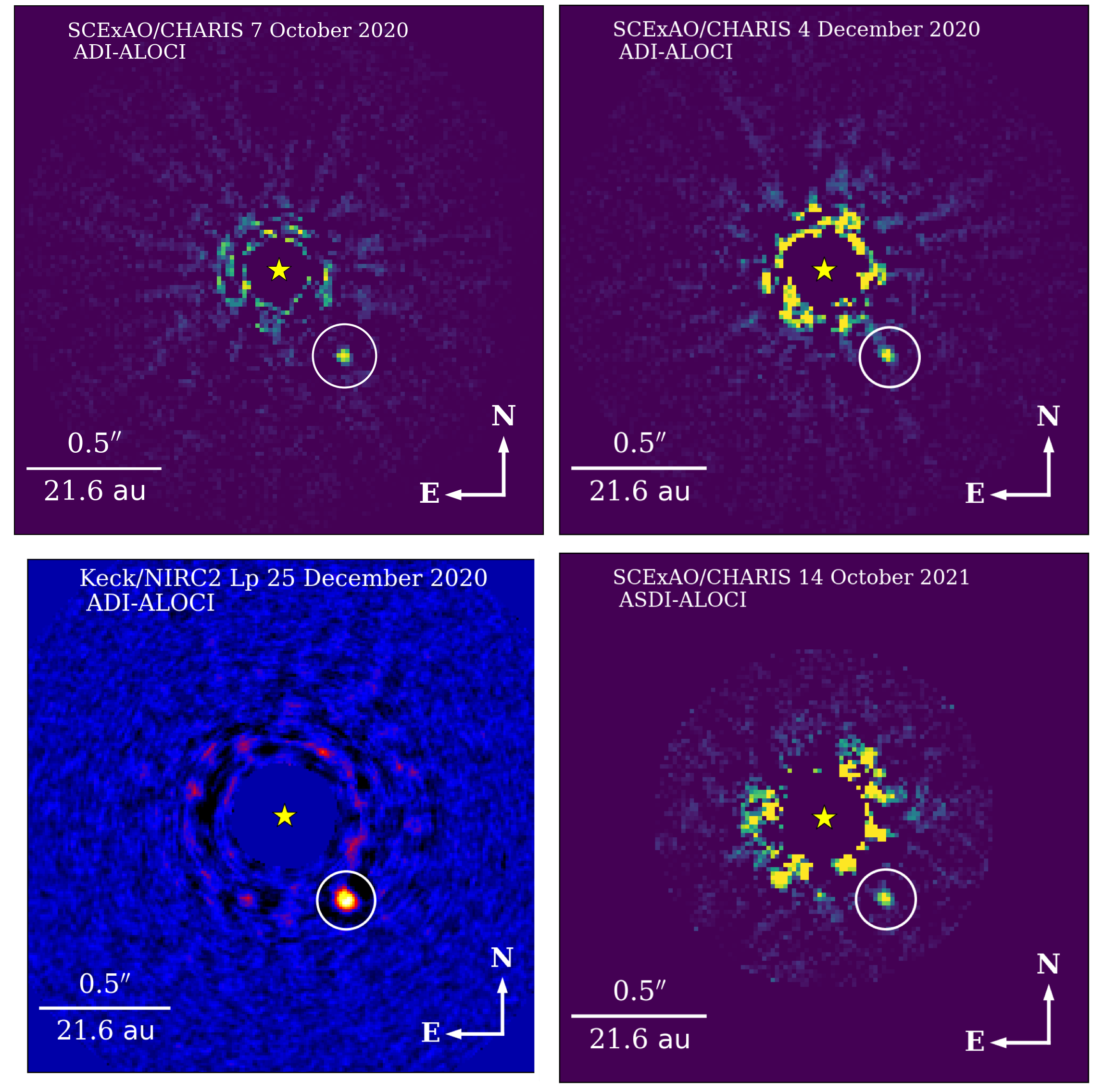}
\caption{
Images of HIP 21152 B (circled) detected from our SCExAO/CHARIS and Keck/NIRC2 data.
The areas close to the central star are masked.
}
\label{fig: gallery}
\end{figure*}

Figure \ref{fig: gallery} shows the HIP 21152 images obtained from our four data sets, from which HIP 21152 B is detected at signal-to-noise ratios\footnote{The SNRs were calculated with the correction of \citet{Mawet2014}.} (SNRs) of 10--19.
We achieve comparable detections with the proprietary version of ALOCI, although the throughput of the proprietary version is far higher.
SDI increases the SNR of the detection at the expense of greater spectroscopic uncertainty.
To correct our spectrophotometry and astrometry for biasing due to processing, we carried out forward modeling as in previous work \citep{Currie2018}.
The forward modeling on the ADI+SDI (ASDI) PSF subtraction accounts for the companion's spectral type.

\subsection{High-Resolution Doppler Spectroscopy}

We monitored HIP 21152A with the high-efficiency fiber-link mode of the HIDES spectrograph equipped on the Okayama 188cm telescope \citep[HIDES-F;][]{Kambe_2013} to measure the star's radial velocities (RVs).
Our monitoring was conducted for about one year from 2011-12-30 and two years from 2020-02-11.
In December 2018, the spectrograph was re-arranged to improve the stability of RV measurements against temperature fluctuations.
We used an image slicer, setting the spectral resolving power to be 55,000 by a 3.8-pixel sampling.
The spectra of HIP 21152 passed through an I$_{2}$ cell, whose absorption features imposed on the spectra are used as references for RV calibration.
Except for three poor-SNR ($<$ 30) spectra, we obtained 32 I$_{2}$-imposed spectra of HIP 21152A with integration times (IT) of 900 or 1800 seconds, and four I$_{2}$-free spectra at various nights (IT = 1800 $\times$ 4 seconds). The I$_{\rm{2}}$-imposed spectra have SNRs ranging from 57 to 258 at $\approx$5500 \AA.\
Our RV calculations adopt a wavelength range from 5028 to 5753 \AA, which contains numerous I$_{2}$ features and little telluric absorption.
The data calibrations and extractions of one-dimensional spectra were performed in a standard way based on IRAF.
The one-dimensional I$_{2}$-free spectra are combined into a single template spectrum after removing outliers, applying 3-pixel median smoothing\footnote{The broadening of absorption lines are limited by the star's rapid rotation even performing 3-pixel smoothing.} and barycentric correction to each spectrum.
The same master template spectrum was compared with each of the 32 I$_{2}$-imposed spectra to measure the RVs of HIP 21152 without producing an offset in the measurements.
Our RV measurements were obtained using the pipeline of \cite{Sato_2002_RVpipeline, Sato_2012_RV_pipleine_II}, which corrects the line profile fluctuations originating from the instrumental instability by modeling them from I$_{2}$ absorption lines.
The spectra were divided into several segments with wavelength widths of $\approx$5.3--6.1 \AA\ and the RVs were calculated in each segment.
The wavelength widths of each segment were set to be much wider than standard RV measurements in HIDES because of HIP 21152's rapid rotation.
The adopted widths provide the smallest RV errors among several attempted segment widths.
The segment-by-segment RVs were statistically summarized to be the final RV measurements in Appendix \ref{sec: RV_measures}.

\begin{figure*}
\centering
\includegraphics[width=1.0\textwidth]{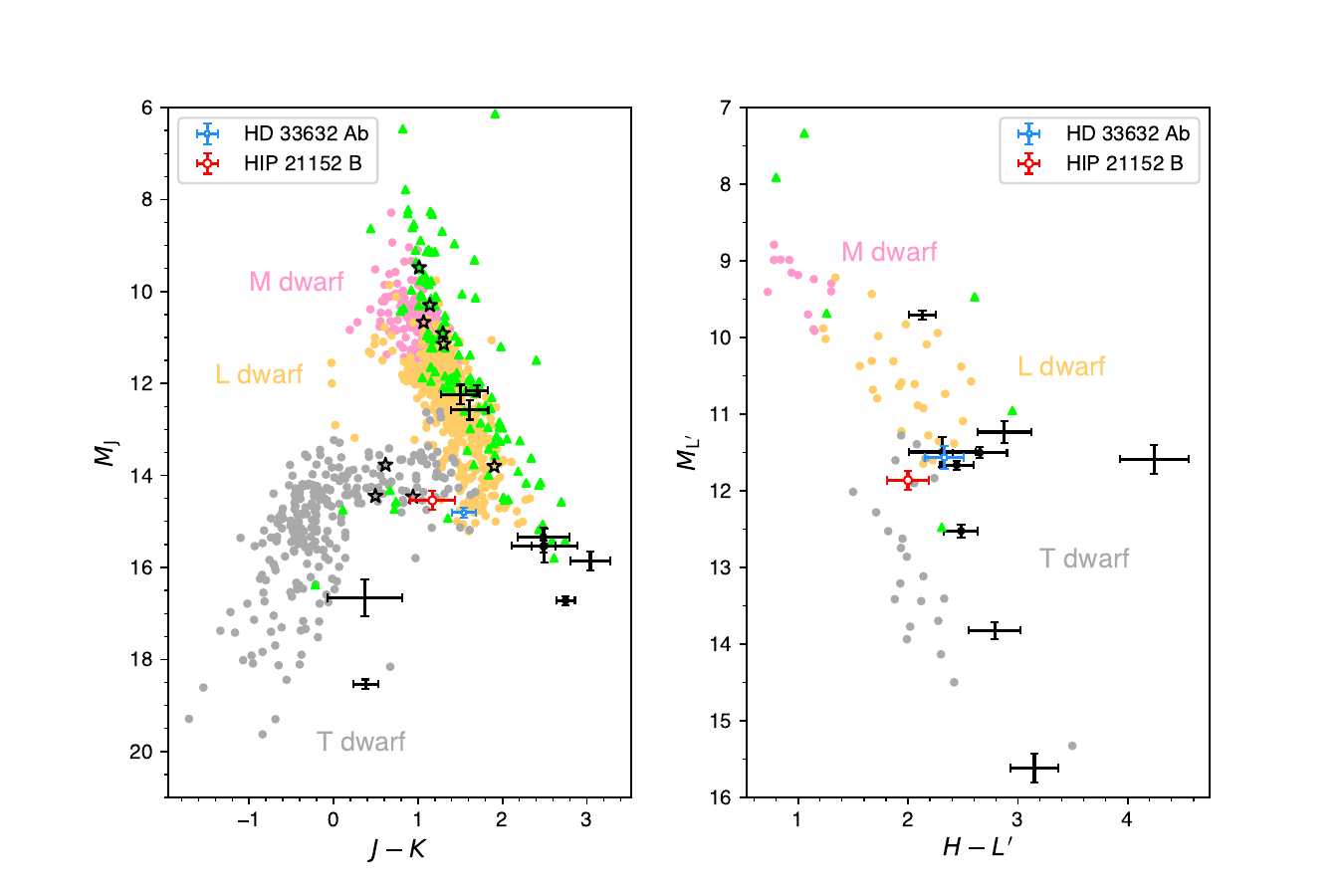}
\vspace{-0.3in}
\caption{
Color-magnitude diagram comparing HIP 21152 B (red open circle) to other substellar objects. (Left) $J - K$ vs. absolute $J$ magnitude ($M_{\rm{J}}$) and (Right) $H - L^{\prime}$ vs. absolute $L^{\prime}$ magnitude ($M_{\rm{L^{\prime}}}$).
Triangle plots correspond to young or low surface-gravity MLT-type dwarfs.
Some of known representative substellar companions are shown by plots with one-$\sigma$ error bars \citep[data are from][]{DupuyLiu2012,Best_2020_ULTRACOOL_DWARF,deRosa2016_HD95086,Kuzuhara2013,Janson_2013_GJ504,Rajan2017,Currie2020}; the planets around HR 8799 are indicated by the filled circles, which are made with the data from \citet{Marois2008a,Metchev_2009_HR8799,Skemer_2014_HR8799,Currie2014_HR8799,Zurlo2016} and the star's 2MASS magnitudes \citep{Skrutskie_2006_2MASS,doi-2MASS}.
The distance modules of the substellar companions are based on \cite{Bailer-Jones_2021_Distance}.
All data of the MLT dwarfs are in the Maunakea Observatories passbands and taken from the compilations in \cite{DupuyLiu2012}, \cite{Leggett_2010_BDMid_phot}, and the database of \cite{Best_2020_ULTRACOOL_DWARF}.
Star symbols represent the Hyades members (Banyan-$\Sigma$ probabilities $>$ 80\%): photometry, parallax, and membership data are from \citet{Best_2020_ULTRACOOL_DWARF, Lodieu_2019_Hyades}.
}
\label{fig: CMD}
\end{figure*}

\section{Infrared Colors, Spectrum, and Atmosphere of HIP 21152 B}
\label{spec_photometry_analysis}

We base the following discussions on the 2020-October spectra (after correcting for spectrophotometry bias) presented in Appendix \ref{app: spectrum_data} because the data at this epoch have the highest SNR, the most stable PSF quality, and the best calibration (see Appendix \ref{app: spec_dec}).
We calculated $J$-, $H$-, and $K_{\rm{s}}$ photometry from the 2020-October ADI spectrum using the bandpasses' filter transmission profiles: $J = 17.72\ \pm\ 0.20 $, $H = 17.04\ \pm\ 0.15$, and $K_{\rm{s}} = 16.55\ \pm\ 0.17$ mag.
From NIRC2 data, we measure $L^{\prime} = 15.04 \pm 0.12$ for HIP 21152~B\footnote{The companion's $L^{\prime}$ photometry was calibrated with HIP 21152A's $L^{\prime}$ magnitude calculated from its 2MASS $K_{\rm s}$-band photometry, the 2MASS color transformations from \citet{Carpenter2001}, and an F5-type star's $K - L$ color \citep[0.04;][]{KenyonHartmann1995}.}.
Figure \ref{fig: CMD} compares the near-infrared colors of HIP 21152~B to those of directly imaged BDs and young exoplanets.
HIP 21152~B's colors are best reproduced by early T dwarfs, near the L-T transition, and are slightly bluer than HD 33632Ab.

\begin{figure*}
\includegraphics[width=1.0\textwidth]{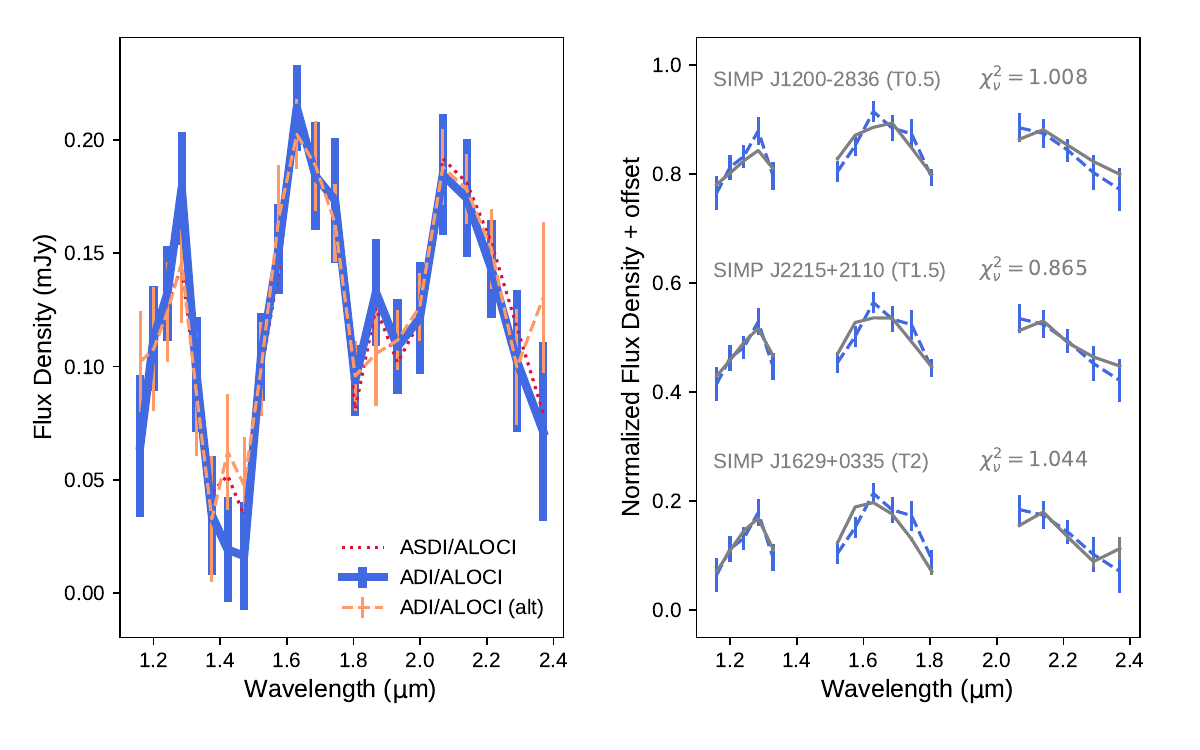}
\vspace{-0.5in}
\caption{
The October 2020 $JHK$ spectra of HIP 21152 B. (Left) The three lines correspond to the fiducial ADI, ASDI, and alternate ADI (labeled as ``alt'') reductions. One-$\sigma$ error bars are appended to the ADI spectra.
(Right) Comparisons between the fiducial spectrum of HIP 21152 B (blue dashed lines) and three template spectra from the Montreal Spectral Library (gray solid lines).
The wavelengths where telluric absorption is significant are masked and were not used in our spectral typing.
The reduced chi-square values ($\chi^2_{\rm{\nu}}$) computed by least-square fitting (15 DOF = sixteen data points minus one optimized parameter for spectrum scaling) are shown above each comparison with the names of the compared BDs and their spectral types.
}
\label{fig: spectrum_type}
\end{figure*}

The extracted spectra of HIP 21152~B are shown in Figure \ref{fig: spectrum_type}, where we plot both the ADI and ASDI spectra as well as the ADI spectrum reduced with a proprietary code; measurements in each spectral channel agree between all reductions.
HIP 21152~B's spectral shape shows strong absorption attributed to water opacity at the gaps between major near-infrared filters and a very blue slope at 2.2--2.4 $\mu$m consistent with methane absorption.

To more quantitatively determine HIP 21152 B's spectral type, we performed a least-square analysis by comparing the spectrum of HIP 21152 B with the template spectra of cool dwarfs \citep{Currie2018}.
The template spectra were taken from the Montreal Spectral Library\footnote{\url{https://jgagneastro.com/the-montreal-spectral-library/}} \citep[e.g.,][]{Gagne2015_Substellar_MG,Robert_2016_dwarf_spec}.
We analyzed the 2020-October spectrum of HIP 21152 B obtained with the ADI-based PSF subtraction.
The fit accounted for the spatially and spectrally correlated noise in an IFS spectrum using the scheme developed by \cite{GrecoBrandt2016}.
We found that 95\% off-diagonal elements of the spectral covariance were smaller than $\approx$0.16 at the companion's angular separation, indicating that the noise is only weakly correlated both spatially and spectrally.
The six wavelength channels affected by significant telluric absorption were omitted in the fit.

The $\chi^{2}$-based comparison shown in Figure \ref{fig: spectrum_type} compares HIP 21152 B's spectrum to selected objects from the Montreal library.
Overall, it is best fit by the T1.5-dwarf SIMP J2215+2210.
Furthermore, the earlier-type template spectrum that best matches HIP 21152~B is the T0.5 object SIMP J1200--2836 ($\Delta \chi^2 \simeq$ 2), while the later-type best match is T2 ($\Delta \chi^2 \simeq$ 2; SIMPJ 1629+0335).
Alternate ADI reductions find similar results.
Earlier L-type dwarfs predict troughs in the water bands bracketing $J$, $H$, and $K$ to be too shallow and have slopes in the 2.2--2.4 $\mu$m range too red to be consistent with HIP 21152 B.
Hence, we estimate the most-likely spectral type of the companion as T1.5$^{+0.5}_{-1.0}$, which is slightly later-type than HD 33632Ab \citep[L9.5$^{+1.0}_{-3.0}$;][]{Currie2020}.
Following \citet{Stephens2009}, HIP 21152 B's spectral type range implies an effective temperature of $T_{\rm eff}$ $\sim$ 1200--1300 K, similar to the 1200--1400 K temperature estimated for HD 33632 Ab.
The relationship between bolometric correction in $H$ band ($BC_{\rm{H}}$) and spectral type from \citet{Liu2010_bck} provides $BC_{\rm{H}}$ = $2.56_{-0.07}^{+0.07}$ for HIP 21152 B.
Assuming a solar bolometric magnitude of 4.74 \citep{Willmer_2018_Sun} and the distance of 43.208$^{+0.050}_{-0.049}$ pc for the HIP 21152 system \citep{Bailer-Jones_2021_Distance}, the companion's bolometric luminosity is then $\log{(L/ L_{\rm{\odot}})}$ = $-$4.673 $\pm$ 0.066.

\section{Astrometric Analysis}

\subsection{HIP 21152 B Astrometry}
We measured the projected separations ($\rho$ in unit of milli-arcseconds or mas) and the position angles (PAs) of HIP 21152 B relative to its central star by fitting elliptical Gaussian models to the companion PSFs identified in all the ALOCI-processed images.
For the CHARIS images, the PSF-fitting was conducted in the images after median-combining all the wavelength channels.
Forward-modeling allowed us to assess astrometric biasing due to processing.
Table \ref{obslog_hip21152} shows the HIP 21152 B astrometry.
The astrometric errors were estimated by taking into account contributions from noise including speckle residuals, calibration errors on the plate scales and true-north angles of our used instruments, and systematic errors in the measurements of the primary star's absolute centers (see Appendix \ref{app: emp_astro_err} for detail and the error budgets).

HIP 21152 B's motion is inconsistent with the relative motion of a background object expected from the astrometry information of the central star from Gaia EDR3 \citep{Gaia_2021_Summary}: the measured vs. predicted position in October 2021 differs by more than 10$\sigma$ in both $\rho$ and PA (see also Appendix \ref{app: cpm_orbit}).

\subsection{Orbit and Dynamical Mass Estimates} \label{sec: orvara}
To constrain HIP 21152 B's orbit and dynamical mass, we model the star's absolute astrometry from HGCA, the star's RV measurements, and the companion's relative astrometry from direct imaging, using the \texttt{orvara} software \citep{Brandt_2021_ORVARA}.
Our \texttt{orvara} analysis carries out Markov Chain Monte Carlo (MCMC) simulations adopting 15 temperatures in the parallel tempering chain and 100 walkers.
Each chain has 7$\times$10$^{5}$ steps; we save every 25th step for 28,000 steps per walker.
We assume a Gaussian prior for the stellar mass: its mean and standard deviation are set to be 1.3 and 0.1 $M_{\rm{\odot}}$ following \cite{David2015}.
For the other parameters, our fit assumed the default priors of \texttt{orvara} including $1/M$ prior for the companion's mass \citep[][see also Appendix \ref{app: cpm_orbit}]{Brandt_2021_ORVARA}.
RV jitter was simulated in the range of 0--100 m s$^{-1}$ with a log-flat prior.
We discarded the initial 2,500 steps as burn-in phase from the recorded 28,000 steps.
With 100 walkers, we have $2.55 \times 10^6$ samples for inference.

\begin{figure*}
     \includegraphics[width=1.0\textwidth]{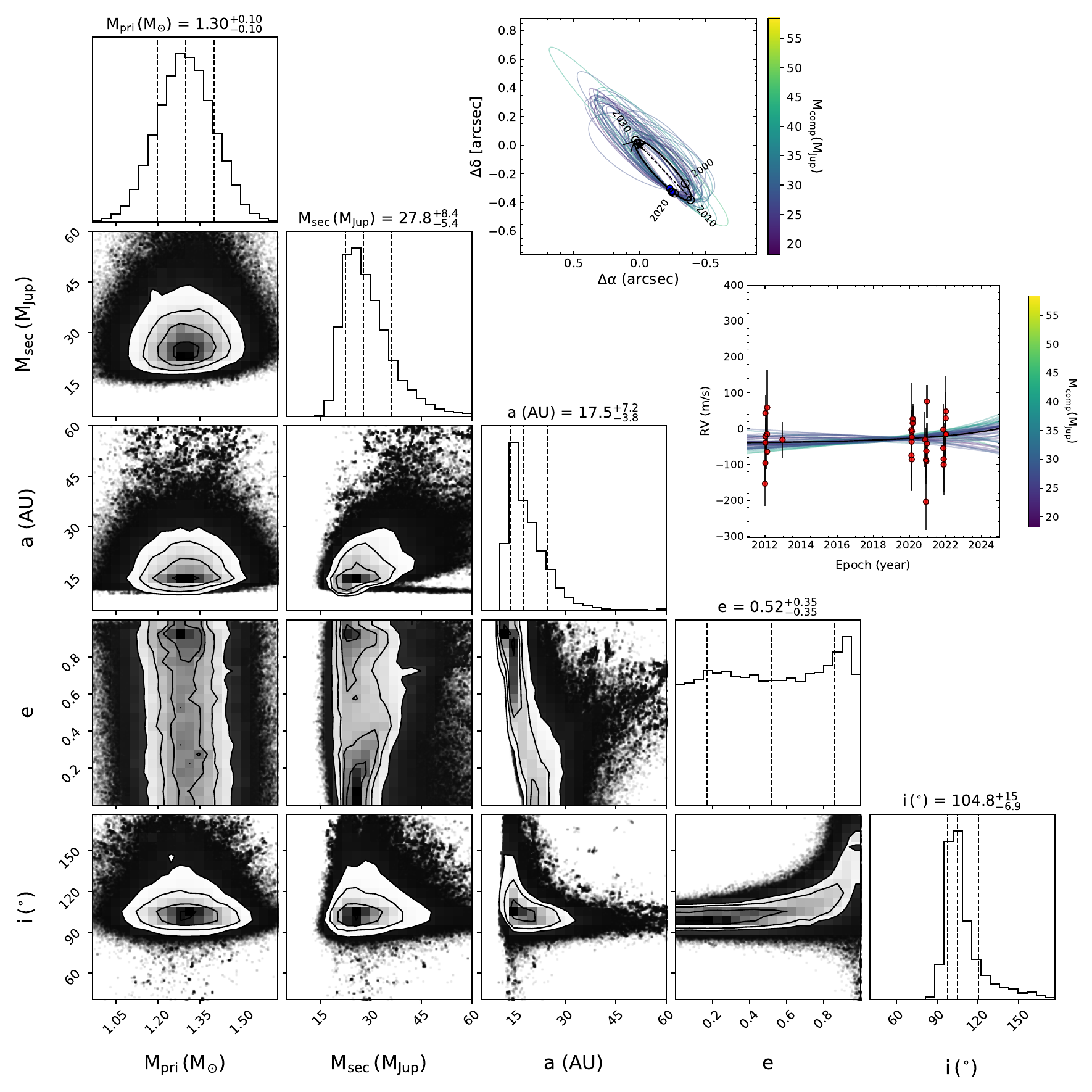} \\
    \caption{Corner plot showing MCMC posterior distributions of host-star mass (M$_{\rm{pri}}$ in unit of $M_{\rm{\odot}}$), companion mass (M$_{\rm{sec}}$ in unit of $M_{\rm{Jup}}$), semi-major axis (a), eccentricity (e), and inclination (i). HIDES radial-velocity (RV) measurements of HIP 21152 A (lower inset) and relative astrometry of HIP 21152 B from CHARIS and NIRC2 (upper inset) are shown, with the best-fit orbit (black solid line) along with 100 orbits randomly taken from our MCMC chains that are color-coded by HIP 21152 B's mass corresponding to the color bars near the inset panels.
    See also Appendix \ref{app: cpm_orbit} for the fitting results to the other measurements.
    }
    \label{fig:orbits}
\end{figure*}

Figure \ref{fig:orbits} shows the corner plot of fitted system parameters from \texttt{orvara}, predicted orbits, and predicted RVs (see Appendix \ref{app: cpm_orbit} for the fits to the other measurements).
The median and the 16--84th percentiles of the MCMC posteriors are provided in Appendix \ref{app: cpm_orbit}.
HIP 21152 B has a best-fit semi-major axis ($a$) of 17.5$^{+7.2}_{-3.8}$ au, viewed at a high inclination of $i$ = 104.8$^{+15}_{-6.9}$\arcdeg.
The estimated mass of the primary largely reflects our input prior, while the companion's mass was estimated to be 27.8$^{+8.4}_{-5.4}$ $M_{\rm Jup}$.
We find no strong constraints on the eccentricity ($e$) of the companion posterior.

\section{Discussion}
We directly imaged a substellar companion orbiting $\sim$18 au from the Sun-like star HIP 21152, which has an accelerating proper motion.
The companion's spectrum is best reproduced by an object near the L/T transition, plausibly an early T dwarf.
The system is a member of the Hyades open cluster (OC) which has a well-constrained age of 750 $\pm$ 100 Myr.
It is notable that there have been the reports of single BDs \citep[e.g.,][]{Lodieu_2019_Hyades} and BD binaries \citep[e.g.,][]{Duchene_2013_Hyades} directly imaged in this OC.
In contrast, there has been no unequivocal confirmation of directly-imaged companions that are less massive than the hydrogen burning limit \citep[e.g.,][]{Fernandes_2019_evomode_ultracool}, around main-sequence stars in Hyades (see a note in Appendix \ref{app: other_bd}).
Accordingly, HIP 21152 B is a crucial benchmark to understand substellar-mass objects as below. \par
HIP 21152 B's dynamical mass is approximately twice the deuterium burning limit and at/slightly above the estimated turnover in mass separating massive jovian exoplanets from BDs \citep{Sahlmann2011}.
Table 1 in \citet{Franson_2022_HD984} lists all the ages and dynamical masses of directly-imaged substellar companions.
We can compare those companions with HIP 21152 B, which has a fractional age uncertainty of 13\% (100 Myr / 750 Myr).
Smaller fractional uncertainties in age estimations lead to smaller fractional uncertainties in mass estimations for directly-imaged substellar companions.
For example, the models of \citet{Baraffe2003} convert a luminosity of $\log{(L/ L_{\rm{\odot}})} \approx -4.9$ to 8 $\pm$ 1 $M_{\rm{Jup}}$ (26 $\pm$ 4 $M_{\rm{Jup}}$) at an age of 50 $\pm$ 10 Myr (500 $\pm$ 100 Myr); meanwhile, the same luminosity corresponds to 8 $\pm$ 2 $M_{\rm{Jup}}$ (26$^{+6}_{-7}$ $M_{\rm{Jup}}$) at 50 $\pm$ 20 Myr (500 $\pm$ 200 Myr).
The corresponding fractional uncertainties for all cases in the list of \citet{Franson_2022_HD984} are larger than 16\%, except for the planets around $\beta$ Pic, which are 13\% just as for HIP 21152 B.
Furthermore, the majority (11/18) of the companions have fractional uncertainties higher than $\sim$30\%.
Thus, the smallest fractional age uncertainty of HIP 21152 B provides the highest fidelity model-dependent mass estimations, besides $\beta$ Pic bc.

In the list, $\beta$ Pic bc and HR 8799 e are the only directly imaged giant planets whose masses have been dynamically constrained.
HIP 21152 B is the closest to those benchmark planets in terms of dynamical mass, best helping unveil the physical and chemical connection between giant planets and BDs.
We note that only HIP 21152 B is firmly associated with an OC among these benchmark substellar companions.
The methods used to characterize stellar and substellar objects such as age estimation techniques \citep[e.g.,][]{Mamajek2008} and evolutionary models \citep[e.g.,][]{Tognelli_2020} have been calibrated by the observations of OCs.
The Hyades OC is especially useful for such calibrations due to its proximity to the Sun.
Hence, HIP 21152 B would be available as one of the best benchmark companions to test evolution and atmosphere theories of cool objects among directly-imaged BDs with inferred dynamical masses. \par

In contrast to the dynamical mass, given HIP 21152 B's bolometric luminosity of $\log{(L/ L_{\rm{\odot}})}$ = $-$4.673 $\pm$ 0.066 and a system age of $\approx$ 650--850 Myr, the \citet{Baraffe2003} luminosity evolution models yield slightly higher predicted masses of 33--42 $M_{\rm Jup}$.
However, temperatures implied by this age range -- 1200--1350 K -- are broadly consistent with spectroscopically-derived values.
Thus, HIP 21152 B may provide another example of 1--2$\sigma$ tension between substellar dynamical masses and those inferred from luminosity evolution at intermediate ages.
\citet{Dupuy2009,Dupuy2014} found significant discrepancies between the dynamical masses and luminosity evolution-inferred masses for the BD binaries GJ 417 BC and HD 130948 BC.
Both systems have ages comparable to Hyades members but have masses of $\approx$ 50 $M_{\rm Jup}$, or $\sim$ 50\% higher than HIP 21152 B.
Future astrometric monitoring of HIP 21152 B will further tests that luminosity evolution models can overestimate the masses of substellar objects.

HIP 21152 B is a benchmark to test atmosphere models of substellar objects as well.
For instance, gravity-sensitive absorption features such as \ion{K}{1} and FeH can be measured via medium-to-high resolution spectroscopy \citep[e.g.,][]{Martin_2017_surfacegravity}.
Those measurements allow the comparison of HIP 21152 B's surface gravity estimated by atmosphere models with that constrained by its dynamical mass and radius \citep[which is appropriately assumed to be about 0.1 $R_{\rm{\odot}}$ at Hyades's age;][]{Baraffe2003}. It is also interesting to characterize HIP 21152 B in the context of L/T transition of substellar objects depending on surface gravity and metallicity \citep[e.g.,][]{Faherty_2012}.
HIP 21152 B benefits such characterization as an anchor point, due to a super-solar metallicity expected from its membership to the Hyades OC \citep{Gagne2018,Gossaage_2018_Hyades} and the semi-empirically constrained surface gravity.

HIP 21152 B's companion-to-primary mass ratio, $q$, is $\sim$ 2.0\%$^{+0.7\%}_{-0.4\%}$. This value is intermediate between bona fide directly imaged exoplanets like HR 8799 bcde \citep[e.g.,][]{Marois2008a,Currie2014_HR8799} and BD companions imaged around Sun-like stars such as HD 33632 Ab \citep{Currie2020} and HD 47127 B \citep{Bowler_2021_HD47127}.
Very few binary star companions have mass ratios this low \citep{Kraus2008}; surveys suggest that the substellar mass function turns over at a mass ratio of $q$ $\sim$ 0.025, where lower (higher) mass ratio companions may be best interpreted as exoplanets (BDs).  \par
In OCs, the gravitational interactions of passing stars can perturb companions on wide orbits and cause ejections in some cases.
\citet{Fujii_Hori_2019} explored the ejection of planets by modeling stellar encounters in OCs via $N$-body simulations.
They found that ejections do not frequently occur in a low-density OC like the Hyades, even in cases where planets orbit their hosts at semi-major axes of 10--100 au.
Their findings should be applicable also to low mass-ratio companions like HIP 21152 B and consistent with this discovery, contributing to verifying such a theory for dynamics of planet/BD companions in OCs.

Finally, this discovery provides further evidence of the promise of using precision astrometry to select direct imaging targets.
Even for a system 750 $\pm$ 100 Myr old, we were able to directly detect a $\approx$20--30 $M_{\rm Jup}$ companion orbiting on solar system scales with high SNRs, demonstrating the capability of extreme AO instruments to detect cooler companions at 10--20 $M_{\rm Jup}$ on the same scale.
A large sample of directly imaged exoplanets and BDs with high quality spectra, dynamical masses, and well-constrained ages will clarify how atmospheres of substellar objects evolve depending on companion mass and how they link to their formation mechanisms.

\input{acknow}

\appendix

\section{Details in High-Contrast Image Processing}\label{app: detail_highcontrast}
We here describe the specific considerations that were taken during our data reductions for high-contrast imaging.
Poor-quality data cubes needed to be removed, since they affect the data reduction procedure.
In order to remove poor Strehl-ratio CHARIS data cubes, we processed only the data with peak-to-halo ratios greater than 10, which are the signal ratios of satellite spot peaks relative to halos of the central star PSFs.
This criterion led us to exclude 8 (3\%) and 69 (35\%) data cubes from observations obtained in October and December of 2020, respectively.
The October 2021 data are split into two sequences, between which the wind-driven halo changed direction: cubes from one sequence are poorly correlated with the other.
For these data, we retain the first sequence since the AO performance and the change in parallactic angle are better (35 data cubes removed).

For all data sets, we spatially filtered the data using a radial-profile subtraction and subtracted the speckle halo using the ALOCI ADI algorithm \citep{Currie2014_HR8799,Currie2018}. For the CHARIS data of October and December 2020, we truncated the set of reference images for each target image based on the correlation between every target-reference pair, selecting the 120 and 100 best-correlated reference images, respectively.
We did not apply this truncation to the processing for the data of October 2021 due to the small number of available exposures.
Other algorithm parameters defining the geometry over which we optimized our reference PSF construction and subtracted this PSF were varied but were generally close to pipeline default values: an optimization area (in units of PSF footprints) of $N_{\rm A}$ = 100 and a rotation gap of 0.5--0.75 full-width-half-maximum of PSF \citep[see e.g.,][]{Pueyo2012}.

To further explore speckle suppression, we considered two additional approaches.
First, for CHARIS data we also applied an SDI reduction on the post-ADI residuals as performed in \citet{Currie2018}, which improves speckle suppression but may introduce less reliable spectral extractions \citep{Pueyo2012}.
Due to the worse observing conditions at October 2021, our analysis relies on SDI reductions for this epoch.
Second, we made alternate ADI reductions that are different from the main procedures, using a proprietary version of ALOCI for which modifications included varying the optimization/subtraction zone geometries and turning on/off a pixel mask over the subtraction zone.
With the adoption of pixel masking, linear-combination coefficients for reference PSF construction are calculated after masking the pixels in PSF subtraction zones.
This technique has been used elsewhere \citep[e.g.,][]{Pueyo2012,Currie2018} including the public ALOCI pipeline to suppress a bias from companion PSFs and significant self-subtractions; indeed, we obtained better throughputs using this technique.
Meanwhile, the proprietary version of ALOCI has the options to adopt several types of optimization zone shape including the standard shape that has been commonly used \citep[e.g.,][]{Pueyo2012} and optimize the zone geometry.

\section{Supplemental Information for Spectroscopy}
\subsection{Spectroscopy from December 2020 Data}
\label{app: spec_dec}
Inspection of the December 2020 data showed issues with spectrophotometric calibration that impeded our ability to extract a spectrum with a quality comparable to that obtained from October 2020 data.
Using our default ALOCI-ADI reduction, we measure broadband photometry of $J = 18.29\ \pm\ 0.42$, $H = 17.15\ \pm\ 0.19$, and $K_{\rm{s}} = 16.16\ \pm\ 0.16$ mag.
While the $H$-band photometry at October and December 2020 is consistent within 1$\sigma$, the $J$- and $K_{\rm{s}}$-band measurements are discrepant at 1.2$\sigma$ and 1.7$\sigma$ levels. As shown in Figure \ref{fig:Oct2020_Dec2021_spec} (left panel), the differences in spectra between the October 2020 and December 2020 ADI reductions are significantly larger than error bars in the shortest and (especially) longest wavelength channels.

Further investigation of this issue identified some partial mitigation measures. The alternate ADI/ALOCI reduction using pixel masking and a different optimization zone geometry yields better agreement (Figure \ref{fig:Oct2020_Dec2021_spec} left panel), suggesting that contamination from residual speckles may be affecting the $K$ band measurements. \par

\begin{figure*}
\includegraphics[width=1.0\textwidth, trim = 0mm 0mm 0mm 0mm,clip]{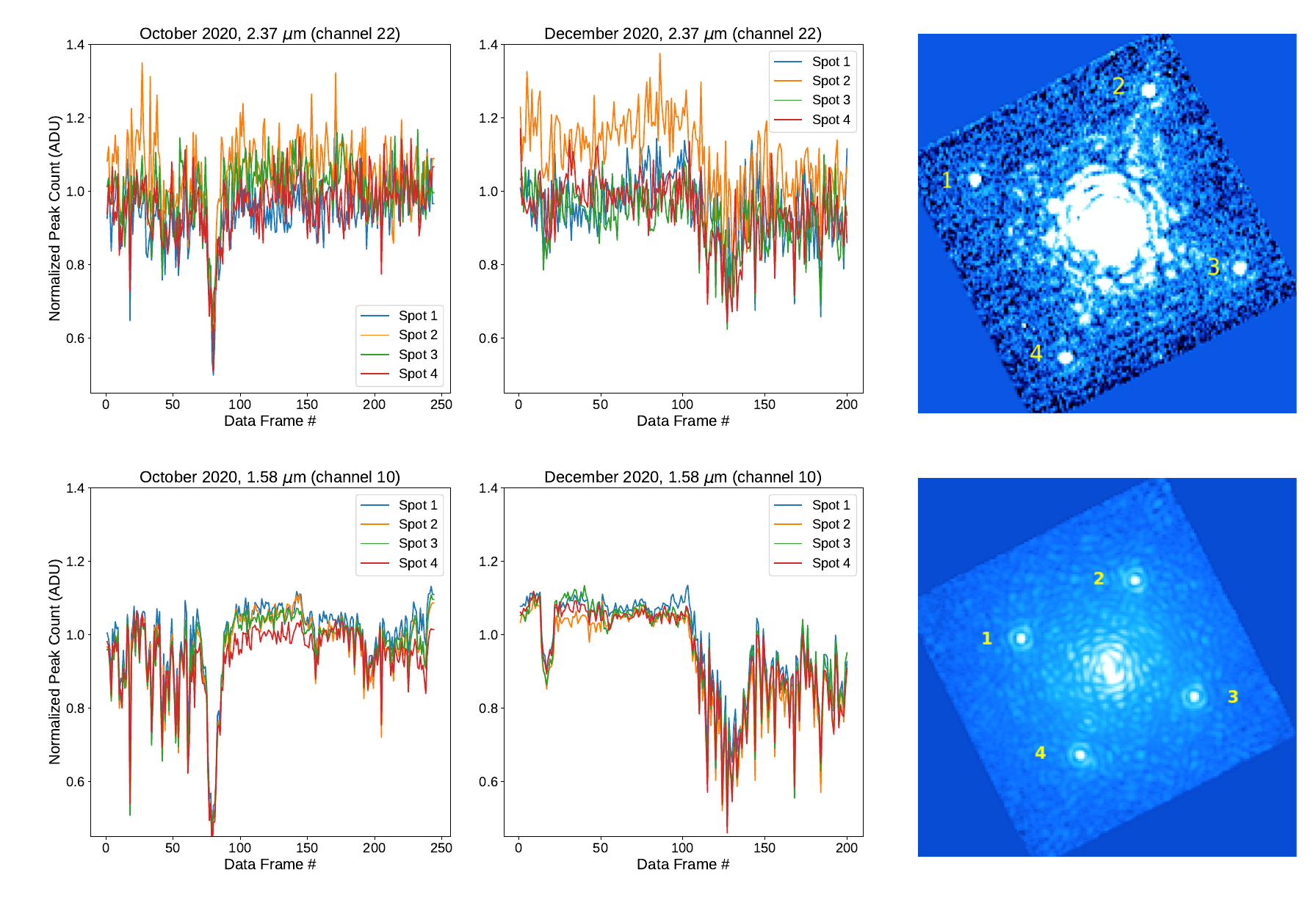}
\caption{Normalized peak count of satellite spots in the
 reddest wavelength (2.37 $\mu m$) and 1.58 $\mu$m slices for cubes in the October 2020 (left) and December 2020 data (middle).   The righthand panels label the satellite spots.
}
\label{fig: deccubes}
\end{figure*}

We also note that the measured spectrum from the December 2020 data can be affected by the ununiform brightnesses of the satellite spots adopted in the spectrophotometric calibration.
The satellite spots in the December 2020 cubes showed modest brightness differences at a channel (2.37 $\mu$m), while they show smaller brightness differences at 1.58 $\mu$m (Figure \ref{fig: deccubes}).
The December 2020 satellite spots show the spot-to-spot ununiformity larger ($\sim$10\%) than the October 2020 spots only in the 2.37 $\mu$m channels (Figure \ref{fig: deccubes}) and telluric-dominated channels, whereas the spots should have roughly equal brightness yielding spectrophotometric precision on the order of $\sim$2\% \citep{Currie2020b}.
\par

Given the apparent problems with the December 2020 data and the higher quality of the October 2020 data, we adopt the October 2020 spectrum as the basis for our analysis\footnote{The October 2021 data are too low in SNR to clarify the true spectrum of HIP 21152 B, though they are consistent with the October 2020 results (see Appendix \ref{app: spec_Oct2021})}.
The lower PSF qualities of December 2020 images are also implied by the larger number of data that we needed to omit (see Section \ref{app: detail_highcontrast}) and the more unstable systematic variations of satellite-spot brightness (see Figure \ref{fig: deccubes}).
We note that the $J-K_{\rm s}$ colors for December 2020 spectra derived from the alternate ADI reduction are $J-K_{\rm s}$ = 1.60 $\pm$ 0.35 with an apparent $J$-band magnitude of 18.05 $\pm$ 0.29 (absolute $J$ band magnitude of 14.87 $\pm$ 0.29).
The resulting color-magnitude diagram positions also lie at the L/T transition as was found for our analyses of the October 2020 spectrum.
Thus, any uncertainties in the shape of HIP 21152 B's spectrum at red wavelengths have a negligible impact on our broad conclusions about HIP 21152 B as a substellar object at the L/T transition.

\subsection{Spectroscopy from October 2021 Data}\label{app: spec_Oct2021}
\begin{figure*}
\includegraphics[width=1.0\textwidth]{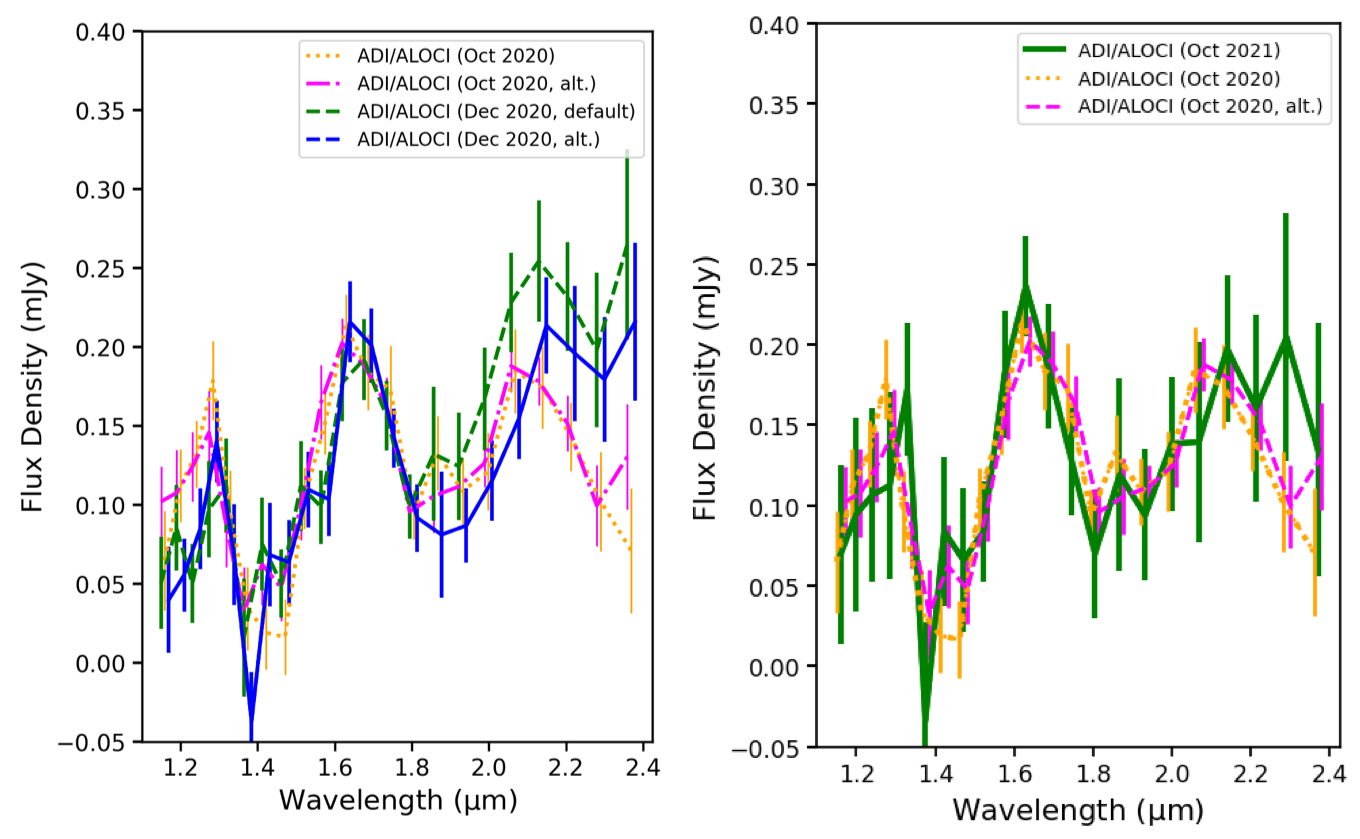}
\caption{
(Left) Comparison between October 2020 spectra and December 2020 spectra extracted using different reduction approaches. (Right)
October 2021 spectrum compared to spectra extracted from October 2020 data. On both the left and right panels, ``alt'' indicates the alternative ALOCI-ADI reductions.
The spectral wavelengths of the spectra on the left and right panels are slightly shifted to avoid plot overlap.
}
\label{fig:Oct2020_Dec2021_spec}
\end{figure*}

The right panels of Figure \ref{fig:Oct2020_Dec2021_spec} compares the spectrum extracted from our October 2021 data set to that taken a year prior.   Overall, the 2021 spectrum is noisier at all wavelengths but agrees in both absolute flux density and shape with the 2020-October results.
New spectra extracted from higher signal-to-noise data and taken at higher resolution are required to advance our understanding of HIP 21152 B's spectral properties.

\subsection{HIP 21152 B Spectrum}\label{app: spectrum_data}
In Table \ref{spectrum_hip21152} below, we present the HIP 21152 B spectrum derived from our standard ALOCI-ADI reduction of the October 2020 data.

\begin{deluxetable}{llll}
     \tablewidth{0pt}
    \tablecaption{HIP 21152 B Spectrum}
    \tablehead{\colhead{Wavelength ($\mu$m)} & \colhead{$F_{\rm \nu}$ (mJy)} &  \colhead{$\sigma$~$F_{\rm \nu}$ (mJy)} & \colhead{SNR}}
    \startdata
1.160 & 0.065 & 0.032 & 2.1\\
1.200 & 0.112 & 0.023 & 5.0\\
1.241 & 0.132 & 0.021 & 6.5\\
1.284 & 0.179 & 0.025 & 8.0\\
1.329 & 0.097 & 0.025 & 3.9\\
1.375 & 0.034 & 0.026 & 1.3\\
1.422 & 0.019 & 0.023 & 0.8\\
1.471 & 0.016 & 0.024 & 0.7\\
1.522 & 0.104 & 0.019 & 5.6\\
1.575 & 0.152 & 0.02 & 8.2\\
1.630 & 0.214 & 0.019 & 12.2\\
1.686 & 0.184 & 0.024 & 9.1\\
1.744 & 0.173 & 0.028 & 7.0\\
1.805 & 0.094 & 0.016 & 6.5\\
1.867 & 0.133 & 0.024 & 5.9\\
1.932 & 0.109 & 0.021 & 5.5\\
1.999 & 0.121 & 0.025 & 5.1\\
2.068 & 0.185 & 0.027 & 8.8\\
2.139 & 0.174 & 0.026 & 7.7\\
2.213 & 0.143 & 0.022 & 7.6\\
2.290 & 0.102 & 0.031 & 3.3\\
2.369 & 0.071 & 0.039 & 1.8\\
\enddata
\tablecomments{Throughput-corrected HIP 21152 B spectrum extracted from 7 October 2020 data, reduced using the ADI/ALOCI pipeline for SCExAO/CHARIS.}
    \label{spectrum_hip21152}
\end{deluxetable}

\section{Radial Velocity Measurements of HIP 21152} \label{sec: RV_measures}
In Table 4, we list individual relative radial-velocity measurements for HIP 21152.
\begin{deluxetable}{lcl}
     \tablewidth{0pt}
    \tablecaption{Relative Radial Velocity (RV) Measurements of HIP 21152 }
    \tablehead{\colhead{JD (days)} & \colhead{RV } &  \colhead{$\sigma_{\textrm{RV}}$}}
    \startdata
2455926.0892653 & $-$121.3 & 62.0\\
2455933.0817933 & $-$63.4 & 51.8\\
2455933.9964362 & 75.7 & 51.9\\
2455934.9652493 & $-$6.2 & 52.9\\
2455936.0063207 & 11.3 & 52.2\\
2455967.0789619 & 17.2 & 58.2\\
2455969.0504358 & $-$31.9 & 48.5\\
2455970.0574695 & 91.4 & 105.8\\
2456281.1998284 & 1.6 & 50.2\\
2458890.9400706 & $-$42.1 & 99.4\\
2458890.9510289 & 28.8 & 132.0\\
2458893.9469028 & $-$4.1 & 53.4\\
2458896.9166423 & 9.1 & 73.8\\
2458896.9275914 & $-$53.6 & 84.9\\
2458899.9304815 & 24.7 & 61.2\\
2458916.9258056 & 59.5 & 41.8\\
2458916.9367650 & 47.9 & 48.3\\
2459166.1754479 & 2.0 & 75.9\\
2459185.0422269 & $-$171.7 & 78.4\\
2459185.0531852 & $-$54.9 & 51.1\\
2459192.9858299 & $-$58.3 & 62.8\\
2459192.9967882 & $-$30.1 & 64.1\\
2459204.0201065 & 108.4 & 46.2\\
2459204.0310602 & $-$8.9 & 56.9\\
2459537.1362429 & 29.8 & 70.7\\
2459537.1468725 & $-$21.9 & 86.4\\
2459543.2202619 & $-$52.5 & 100.8\\
2459543.2313550 & $-$68.0 & 75.5\\
2459584.1654891 & 61.7 & 118.2\\
2459586.0721793 & 16.2 & 79.2\\
2459588.0634205 & 17.0 & 59.4\\
2459588.0744923 & 80.5 & 54.7\\
 \enddata
    \tablecomments{The measured RVs and their errors are given in unit of  m s$^{-1}$.
    }
\label{tab: rv_hip21152}

\end{deluxetable}

\section{Supplemental Information for Astrometric Analysis}\label{app:astromerror}
\subsection{Empirical Analysis of Relative Astrometry Measurements} \label{app: emp_astro_err}
As CHARIS was craned in and out of position multiple times between our observations, we reassessed the plate scale and true-north orientation angle offset of the CHARIS detector.
As in \citet{Currie2018,Currie2020}, we used SCExAO/CHARIS and Keck-II/NIRC2 data of the companion around HD 1160 to identify any change in detector astrometric properties\footnote{This analysis is described in full in an upcoming paper (Torres-Quijado, Currie, et al., in prep.)}.
As described in \cite{Currie_2022_ABAur}, these analyses favor a slightly revised plate scale of 16.15 $\pm$ 0.05 mas pixel$^{-1}$ but otherwise no measureable changes in the astrometric calibration determined in \citet{Currie2018}.
Our analysis is based on the revised pixel scale described above, and the true-north orientation offset angle in \citet{Currie2018}.\par
Contemporaneous CHARIS and NIRC2 astrometry for the HD 1160's companion find consistent results: e.g., $\rho$ = 0\farcs{}791 and 0\farcs{}797; positiona angle (PA) = 244.60\arcdeg\ and 244.80\arcdeg\ for high quality October 2020 and lower-quality December 2020 CHARIS data compared to $\rho$ = 0\farcs{}791, PA = 244.79\arcdeg\ for high-quality NIRC2 data.
Our NIRC2 astrometric measurements are based on the distortion calibration of \cite{Service2016}, from which we adopt the plate scale of 9.971 $\pm$ 0.004 mas pixel$^{-1}$ and the true-north orientation offset of 0.262\arcdeg\ $\pm$ 0.020\arcdeg.

We also consider an absolute astrometric error due to uncertainties in determining the star's positions.
For the Keck/NIRC2 data, \citet{Konopacky_2016_HR8799} quote an uncertainty of $\sim$ 2 mas in the star's center when determined through the partially transmissive Lyot coronagraph.
For CHARIS, internal source tests described in \citet{Currie2020b} included analyses of the star position determined from fitting the satellite spots vs. unobstructed PSF centroids.    The tests reveal up to a 0.25 pixel offset ($\sim$4 mas) in reported position of the centroid estimated from satellite spots and that determined from an unobstructed PSF.
The NKT Photonics SuperK laser we used for this analysis only extends to 1.7 microns, so we do not have a direct measurement of any biasing at longer wavelengths.   The source of this difference is unclear but could be due to residual field distortion.  As an empirical test, we compared the centroid positions for HD 1160's companion in 7 different data sets from August 2020 to January 2022, a timeframe over which we expect the orbital motion to be negligible; the standard deviation in the east and north positions are $\sim$3.7 mas and $\sim$2.3 mas.
To be conservative, we adopt an absolute astrometric error in each coordinate of 0.25 pixels (= 4 mas) for CHARIS. \par

For CHARIS, uncertainties from astrometric biasing due to processing and (for the October and December 2020 data sets) the intrinsic detection SNR are small compared to intrinsic uncertainties in the pixel scale (0.05 mas), north position angle offset (0.27\arcdeg), and absolute centroid measurement ($\sim$ 4 mas).   However, residual and only partially-whitened speckle noise may contaminate centroid measurements more than expected from an SNR estimate \citep[e.g. as in ][]{Gaspar2020}.
We simulated noise-injected companion PSFs to estimate the random uncertainties of the companion centroids in the CHARIS images, providing an empirically-motivated estimate of our centroid uncertainties.
The simulated PSFs were made by adding noise floors to the forward-modeled PSFs \citep[see Section \ref{sec: obs_reduction} and ][]{Currie2018}.
The noise floors were taken from the areas in the final images created by combining the PSF-subtracted cube frames.
Then, we used the areas at the same separations as the companion but the different 12 PAs, which starts from the companion's PA + 45\arcdeg\ and ends at the companion's PA + 320\arcdeg\ with intervals of 25\arcdeg.
The standard deviations of the centroids calculated from the simulated PSFs were taken to be the random uncertainties of the companion centroid measurements.
These uncertainties [$\sigma_{\rm{x}}$, $\sigma_{\rm{y}}$] are equal to [0.074, 0.109], [0.133, 0.081], and [0.18, 0.17] pixels (= [1.2, 1.8], [2.1, 1.3], and [2.9, 2.8] mas with a plate scale of 16.15 mas pixel$^{-1}$) for the 2020-October, 2020-December, and 2021-October CHARIS images.
We do not perform the same analysis for the NIRC2 $L^{\rm \prime}$ data since the SNR is lower and intrinsic PSF is roughly twice as large as CHARIS's ($\theta$ $\approx$ 0\farcs{}08 vs. 0\farcs{}043).
In this case, the centroid uncertainty estimated from the intrinsic SNR is $\sim$ 0.35 pixels: significantly larger than for any CHARIS measurement.

Another source of the random errors is attributed to the alignment of the individual images.
We evaluated this error source using the SCExAO/CHARIS data sets obtained in October and December 2020 for HIP 21152.
The residuals between the central star's positions calculated at each wavelength channel of a cube and the polynomial function fit to the positions can correspond to the image alignment errors.
The residual scatter of an image alignment is much smaller than the other insignificant error sources (see above), and further decreases when integrating all frames and all channels; we thus neglect the image alignment errors in SCExAO/CHARIS.
Table \ref{tab:relastro_error_detail} summarizes the error evaluations described above for the CHARIS measurements.
When the astrometric calibrations for SCExAO/CHARIS will be updated in future (e.g., for distortion calibration), we recommend the reader to refer to Table \ref{tab:relastro_error_detail} for recalculating the astrometric measurements with the updated calibrations.

\subsection{Common Proper Motion and Orbit Analysis} \label{app: cpm_orbit}

\begin{figure*}
\includegraphics[trim=0mm 0mm 8mm 0mm,clip]{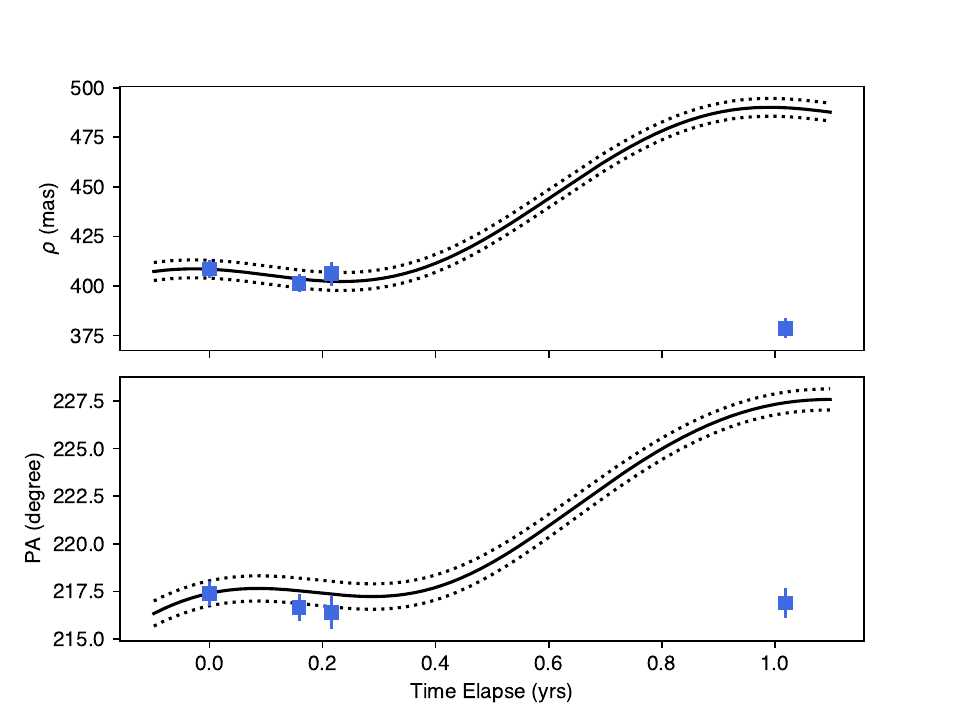}
\caption{Common proper motion analysis for HIP 21152 B.
The projected separation ($\rho$) and position angle (PA) measurements for HIP 21152 B are shown at the top and bottom panels, respectively.
The horizontal axes indicate the time elapsed from the October 2020 epoch. An expected motion assuming HIP 21152 B is a background star is shown by the dashed lines encompassed by their one-$\sigma$ errors.
It is clearly demonstrated that HIP 21152 B cannot be a background star given the large difference between the expected background motion and the measured $\rho$ and PA at the latest epoch (October 2021): 16$\sigma$ in $\rho$ and 11$\sigma$ in PA.
}
\label{fig: propmot}
\end{figure*}

\begin{figure*}
\includegraphics[width=1.0\textwidth,trim=0mm 0mm 0mm 0mm,clip]{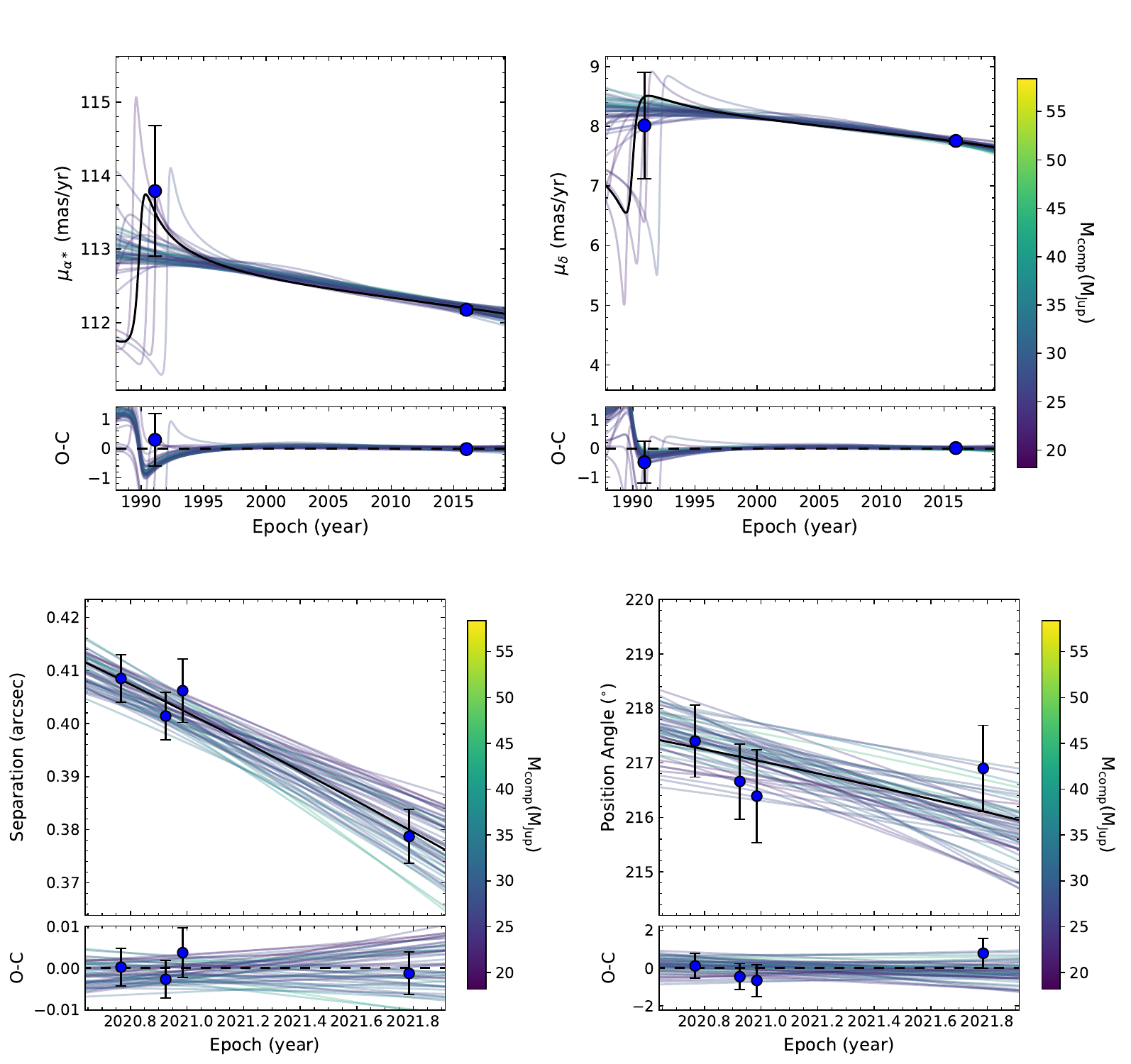}
\caption{HIP 21152A's proper motion variations along RA (top left) and DEC (top right) and the measured projected separations (bottom left) and position angles (PAs; bottom right) of HIP 21152 B relative to HIP 21152 A. The best-fit orbit is indicated by black solid curves, while the randomly-selected 100 orbits are shown by color-coded curves. The color bars near each panel correspond to the companion's mass.
}
\label{fig: orvara_other}
\end{figure*}

In Figure \ref{fig: propmot}, we compare the measured positions of HIP 21152 B relative to HIP 21152A with the positions expected if HIP 21152 B is an unbound background object (i.e., common proper motion analysis).
The expectation was made with HIP 21152A's right ascension (RA), declination (DEC), RA and DEC proper motions, and parallax from Gaia EDR3 \citep{Gaia_2021_Summary}.
The results from our \texttt{orvara} orbit modeling for the HIP 21152 system (Section \ref{sec: orvara}) are summarized in Table \ref{tab:mcmc_results}.
In addition to Figure \ref{fig:orbits}, Figure \ref{fig: orvara_other} shows the fitted orbits to HIP 21152A's proper motion variations from HGCA \citep{Brandt2021_HGCA}
along RA and DEC and HIP 21152 B's projected separations and position angles.

\section{Substellar Companion Candidates around Main-Sequence Stars in the Hyades} \label{app: other_bd}
There have been no previous reports of \textit{confirmed} substellar companions to main-sequence stars in the Hyades open cluster (OC), despite its proximity to the solar system.
We note that \citet{Morzinski_2012} reported candidates of Hyades substellar companions detected with shallow adaptive optics imaging.
In addition, an L1 $\pm$ 1 type companion around the M4-type star 1RXS J034231.8+121622 was reported by \citet{Bowler_2015_PALMS}.
A similarity of the space velocities and the sky position of this system with those of the Hyades OC was already discussed in \citet{Bowler_2015_PALMS}; nevertheless, they concluded that this system's membership to the Hyades OC is unclear.
For 1RXS J034231.8+121622, we run the Banyan-$\Sigma$ algorithm \citep{Gagne2018} with the star's distance adopted in \citet{Bowler_2015_PALMS}, RA, DEC, and proper-motion measurements from \citet{Gaia_2021_Summary}, and absolute RV from \citet{Shkolnik_2012}, providing a zero probability as a Hyades member.
In contrast, we compute a membership probability for the same star of 99.9\% using the same algorithm after updating the distance to a Gaia-based measurement from \citet{Bailer-Jones_2021_Distance}; the different membership probability is therefore caused by Gaia data becoming available.
Although 1RXS J034231.8+121622 B can be identified as a high-probability member of the Hyades OC, it is still unclear whether the companion has a substellar mass.
We calculate a mass of 1RXS J034231.8+121622 B with the new membership and the Gaia-based distance measurement.
For this companion, we adopt a distance of 32.96$^{+0.022}_{-0.024}$ pc from \citet{Bailer-Jones_2021_Distance} and calculate an $H$-band bolometric correction of 2.70 $\pm$ 0.08 mag following \citet{Liu2010_bck} to convert an $H$-band apparent magnitude of 13.51 $\pm$ 0.05 \citep{Bowler_2015_PALMS} to the bolometric luminosity of $\log{(L/L_{\rm{\odot}})} = -$3.552 $\pm$ 0.038.
With the age of Hyades (750 $\pm$ 100 Myr) and the evolutionary models of \citet{Baraffe2003}, the bolometric luminosity is converted to a mass of 76--83 $M_{\rm{Jup}}$, which is near or slightly above the hydrogen burning limit \citep[$\approx$70--80 $M_{\rm{Jup}}$ in general; e.g.,][]{Fernandes_2019_evomode_ultracool}.
We thus find that 1RXS J034231.8+121622 B is a candidate substellar companion in the Hyades OC.

\begin{deluxetable*}{llcccc}
     \tablewidth{0pt}
    \tablecaption{HIP 21152 B relative astrometry uncertainties in CHARIS without systematic errors of 0.25 pixels (see Appendix \ref{app: emp_astro_err}). \label{tab:relastro_error_detail}}
    \tablehead{
    \colhead{Date} &
    \colhead{Instrument} &
    \colhead{[$\sigma_{\rm{x}}$, $\sigma_{\rm{y}}$]} &
    \colhead{$\sigma_{\rm{\rho}}$} &
    \colhead{$\sigma_{\rm{PA}}$}
    \\
    \colhead{(UT)} & \colhead{} & \colhead{(pixels)}  & \colhead{(mas)} & \colhead{(\arcdeg)}
    }
    \startdata
    2020-10-07 & SCExAO/CHARIS & [0.074, 0.109] & 2.0 & 0.34\\
    2020-12-04 & SCExAO/CHARIS & [0.133, 0.081] & 2.1 & 0.38\\
    2021-10-14 & SCExAO/CHARIS & [0.18, 0.17] & 3.0 & 0.51\\
    \enddata
 \tablecomments{Angular separations are in units of milli-arcseconds (mas). The calibration errors in the CHARIS plate scale and true-north orientation offset are included in $\sigma_{\rm{\rho}}$ and $\sigma_{\rm{PA}}$.}
\end{deluxetable*}

\input{mcmc_results_new}

\end{document}

%% file: authors.tex
\correspondingauthor{Masayuki Kuzuhara}
\email{m.kuzuhara@nao.ac.jp}
\author[0000-0002-4677-9182]{Masayuki Kuzuhara}
\affiliation{Astrobiology Center of NINS, 2-21-1, Osawa, Mitaka, Tokyo, 181-8588, Japan}
\affiliation{National Astronomical Observatory of Japan, 2-21-2, Osawa, Mitaka, Tokyo 181-8588, Japan}
\author[0000-0002-7405-3119]{Thayne Currie}
\affiliation{Subaru Telescope, National Astronomical Observatory of Japan, 650 North A`oh$\bar{o}$k$\bar{u}$ Place, Hilo, HI  96720, USA}
\affiliation{NASA-Ames Research Center, Moffett Blvd., Moffett Field, CA, USA}
\affiliation{Eureka Scientific, 2452 Delmer Street Suite 100, Oakland, CA, USA}
\author{Takuya Takarada}
\affiliation{Astrobiology Center of NINS, 2-21-1, Osawa, Mitaka, Tokyo, 181-8588, Japan}
\affiliation{National Astronomical Observatory of Japan, 2-21-2, Osawa, Mitaka, Tokyo 181-8588, Japan}
\author[0000-0003-2630-8073]{Timothy D. Brandt}
\affiliation{Department of Physics, University of California, Santa Barbara, Santa Barbara, California, USA}
\author{Bun'ei Sato}
\affiliation{Department of Earth and Planetary Sciences, School of Science, Tokyo Institute of Technology, 2-12-1 Ookayama, Meguro-ku, Tokyo 152-8551, Japan}
\author[0000-0002-6879-3030]{Taichi Uyama}
\affiliation{Infrared Processing and Analysis Center, California Institute of Technology, Pasadena, CA 91125, USA}
\author[0000-0001-8345-593X]{Markus Janson}
\affiliation{Department of Astronomy, Stockholm University, Stockholm, Sweden}
\author[0000-0001-6305-7272]{Jeffrey Chilcote}
\affiliation{Department of Physics, University of Notre Dame, Notre Dame, IN, USA}
\author[0000-0001-8103-5499]{Taylor Tobin}
\affiliation{Department of Physics, University of Notre Dame, Notre Dame, IN, USA}
\author[0000-0002-6964-8732]{Kellen Lawson}
\affiliation{Homer S Dodge Department of Physics and Astronomy, University of Oklahoma, Norman, OK, USA}
\author[0000-0003-4676-0251]{Yasunori Hori}
\affiliation{Astrobiology Center of NINS, 2-21-1, Osawa, Mitaka, Tokyo, 181-8588, Japan}
\affiliation{National Astronomical Observatory of Japan, 2-21-2, Osawa, Mitaka, Tokyo 181-8588, Japan}
\affiliation{Department of Astronomical Science, The Graduate University for Advanced Studies, SOKENDAI, 2-21-1 Osawa, Mitaka, Tokyo 181-8588, Japan}
\author[0000-0002-1097-9908]{Olivier Guyon}
\affiliation{Subaru Telescope, National Astronomical Observatory of Japan,
650 North A`oh$\bar{o}$k$\bar{u}$ Place, Hilo, HI  96720, USA}
\affil{Steward Observatory, The University of Arizona, Tucson, AZ 85721, USA}
\affil{College of Optical Sciences, University of Arizona, Tucson, AZ 85721, USA}
\affil{Astrobiology Center of NINS, 2-21-1, Osawa, Mitaka, Tokyo, 181-8588, Japan}
\author[0000-0001-5978-3247]{Tyler D. Groff}
\affiliation{NASA-Goddard Space Flight Center, Greenbelt, MD, USA}
\author[0000-0002-3047-1845]{Julien Lozi}
\affiliation{Subaru Telescope, National Astronomical Observatory of Japan,
650 North A`oh$\bar{o}$k$\bar{u}$ Place, Hilo, HI  96720, USA}
\author[0000-0003-4018-2569]{Sebastien Vievard}
\affiliation{Subaru Telescope, National Astronomical Observatory of Japan,
650 North A`oh$\bar{o}$k$\bar{u}$ Place, Hilo, HI  96720, USA}
\author[0000-0003-2806-1254]{Ananya Sahoo}
\affiliation{Subaru Telescope, National Astronomical Observatory of Japan,
650 North A`oh$\bar{o}$k$\bar{u}$ Place, Hilo, HI  96720, USA}
\author[0000-0003-4514-7906]{Vincent Deo}
\affiliation{Subaru Telescope, National Astronomical Observatory of Japan,
650 North A`oh$\bar{o}$k$\bar{u}$ Place, Hilo, HI  96720, USA}
\author[0000-0001-5213-6207]{Nemanja Jovanovic}
\affiliation{Department of Astronomy, California Institute of Technology, 1200 E. California Blvd.,Pasadena, CA, 91125, USA}
\author[0000-0002-1094-852X]{Kyohoon Ahn}
\affiliation{Subaru Telescope, National Astronomical Observatory of Japan,
650 North A`oh$\bar{o}$k$\bar{u}$ Place, Hilo, HI  96720, USA}
\author[0000-0003-1180-4138]{Frantz Martinache}
\affiliation{Universit\'{e} C\^{o}te d'Azur, Observatoire de la C\^{o}te d'Azur, CNRS, Laboratoire Lagrange, France}
\author[0000-0002-9372-5056]{Nour Skaf}
\affiliation{Subaru Telescope, National Astronomical Observatory of Japan, 650 North A`oh$\bar{o}$k$\bar{u}$ Place, Hilo, HI  96720, USA}
\affiliation{LESIA, Observatoire de Paris, Univ.~PSL, CNRS, Sorbonne Univ., Univ.~de Paris, 5 pl.
Jules Janssen, 92195 Meudon, France}
\author{Eiji Akiyama}
\affiliation{Division of Fundamental Education and Liberal Arts, Department of Engineering, Niigata Institute of Technology
1719 Fujihashi, Kashiwazaki, Niigata 945-1195, Japan}
\author[0000-0002-8352-7515]{Barnaby R. Norris}
\affiliation{Sydney Institute for Astronomy, School of Physics, University of Sydney, Sydney, New South Wales, Australia}
\author{Micka{\"e}l Bonnefoy}
\affiliation{Institut de Planétologie et d'Astrophysique de Grenoble / CNRS (IPAG, France)}
\author[0000-0002-7650-3603]{Krzysztof G. He{\l}miniak}
\affiliation{Nicolaus Copernicus Astronomical Center of the Polish Academy of Sciences, ul. Rabia{\'n}ska 8, 87-100, Toru{\'n}, Poland}
\author{Tomoyuki Kudo}
\affiliation{Subaru Telescope, National Astronomical Observatory of Japan, 650 North A`oh$\bar{o}$k$\bar{u}$ Place, Hilo, HI  96720, USA}
\author[0000-0003-0241-8956]{Michael W. McElwain}
\affiliation{NASA-Goddard Space Flight Center, Greenbelt, MD, USA}
\author[0000-0001-9992-4067]{Matthias Samland}
\affiliation{Max-Planck-Institut f{\" u}r Astronomie, K{\" o}nigstuhl 17 D-69117, Heidelberg, Germany}
\author[0000-0002-4309-6343]{Kevin Wagner}
\affiliation{Steward Observatory, The University of Arizona, Tucson, AZ 85721, USA}
\affiliation{NASA Hubble Fellowship Program - Sagan Fellow}
\author[0000-0001-9209-1808]{John Wisniewski}
\affiliation{Department of Physics and Astronomy, University of Oklahoma, Norman, OK, USA}
\author[0000-0002-9259-1164]{Gillian R. Knapp}
\affiliation{Department of Astrophysical Science, Princeton University, Peyton Hall, Ivy Lane, Princeton, NJ 08544, USA}
\author[0000-0003-2815-7774]{Jungmi Kwon}
\affiliation{Department of Astronomy, Graduate School of Science, The University of Tokyo, 7-3-1, Hongo, Bunkyo-ku, Tokyo, 113-0033, Japan}
\author{Jun Nishikawa}
\affiliation{National Astronomical Observatory of Japan, 2-21-1 Osawa, Mitaka, Tokyo 181-8588, Japan}
\affiliation{Astrobiology Center of NINS, 2-21-1, Osawa, Mitaka, Tokyo, 181-8588, Japan}
\affiliation{Department of Astronomical Science, The Graduate University for Advanced Studies, SOKENDAI, 2-21-1 Osawa, Mitaka, Tokyo 181-8588, Japan}
\author{Eugene Serabyn}
\affiliation{Jet Propulsion Laboratory, California Institute of Technology, 4800 Oak Grove Drive, Pasadena, CA 91109, USA}
\author{Masahiko Hayashi}
\affiliation{National Astronomical Observatory of Japan, 2-21-2, Osawa, Mitaka, Tokyo 181-8588, Japan}
\author[0000-0002-6510-0681]{Motohide Tamura}
\affiliation{Astrobiology Center of NINS, 2-21-1, Osawa, Mitaka, Tokyo, 181-8588, Japan}
\affiliation{National Astronomical Observatory of Japan, 2-21-2, Osawa, Mitaka, Tokyo 181-8588, Japan}
\affiliation{Department of Astronomy, Graduate School of Science, The University of Tokyo, 7-3-1, Hongo, Bunkyo-ku, Tokyo, 113-0033, Japan}

%% file: acknow.tex
\acknowledgments
\indent We thank Zhoujian Zhang for discussions about bolometric correction, Michiko Fujii for stellar encounter in open clusters, and Brendan Bowler for the companion of 1RXS J034231.8+121622.
This work has made use of data from the European Space Agency (ESA) mission {\it Gaia} (\url{https://www.cosmos.esa.int/gaia}), processed by the {\it Gaia} Data Processing and Analysis Consortium (DPAC, \url{https://www.cosmos.esa.int/web/gaia/dpac/consortium}). Funding for the DPAC has been provided by national institutions, in particular the institutions participating in the {\it Gaia} Multilateral Agreement.
This research has made use of the Keck Observatory Archive \citep[KOA;][]{doi-KOA}, which is operated by the W. M. Keck Observatory and the NASA Exoplanet Science Institute (NExScI), under contract with the National Aeronautics and Space Administration.
This work has benefited from The UltracoolSheet, maintained by Will Best, Trent Dupuy, Michael Liu, Rob Siverd, and Zhoujian Zhang, and developed from compilations by \citet{DupuyLiu2012}, \citet{Dupuy_Kraus_2013}, \citet{Liu_2016_parallax}, \citet{Best_2018_parallax}, and \citet{Best_2021_parallax}.
This research has benefitted from the Montreal Brown Dwarf and Exoplanet Spectral Library, maintained by Jonathan Gagn{\' e}. 
This research has made use of the SIMBAD and VizieR services, both operated at Centre de Donn{\' e}es astronomiques de Strasbourg (CDS, \url{https://cds.u-strasbg.fr/}) in France, and NASA's Astrophysics Data System Bibliographic Services.
IRAF is distributed by the National Optical Astronomy Observatories, which is operated by the Association of Universities for Research in Astronomy, Inc. (AURA) under cooperative agreement with the National Science Foundation.
Part of data analysis was carried out on the Multi-wavelength Data Analysis System operated by the Astronomy Data Center (ADC), National Astronomical Observatory of Japan. \par
We are honored and grateful for the opportunity of observing the Universe from Maunakea, which has the cultural, historical, and natural significance in Hawaii.
We appreciate the critical support from all the current and recent Subaru and Keck Observatory staffs. 
Their support was essential in achieving this discovery, especially amidst the many difficulties associated with the COVID-19 pandemic.\\
\indent We thank the Subaru and NASA Keck Time Allocation Committees for their generous support of this program.  
TC was supported by a NASA Senior Postdoctoral Fellowship and NASA/Keck grant LK-2663-948181.   
TB gratefully acknowledges support from the Heising-Simons foundation and from NASA under grant \#80NSSC18K0439. 
VD and NS were supported by the NASA grant \#80NSSC19K0336.
KW acknowledges support from NASA through the NASA Hubble Fellowship grant HST-HF2-51472.001-A awarded by the Space Telescope Science Institute, which is operated by the Association of Universities for Research in Astronomy, Incorporated, under NASA contract NAS5-26555.
Part of this work was carried out at the Jet Propulsion Laboratory, California Institute of Technology, under contract with NASA.
The results reported herein benefited from collaborations and/or information exchange within NASA's Nexus for Exoplanet System Science (NExSS) research coordination network sponsored by NASA's Science Mission Directorate.
MT is supported by JSPS KAKENHI Grant \# 18H05442 and EA is supported by MEXT/JSPS KAKENHI grant \# 17K05399.
\\
\indent The development of SCExAO was supported by JSPS (Grant-in-Aid for Research \#23340051, \#26220704 \& \#23103002), Astrobiology Center of NINS, the Mt Cuba Foundation, and the director's contingency fund at Subaru Telescope.  CHARIS was developed under the support by the Grant-in-Aid for Scientific Research on Innovative Areas \#2302.  
The Okayama 188cm telescope is operated by a consortium led by
Exoplanet Observation Research Center, Tokyo Institute of Technology
(Tokyo Tech), under the framework of tripartite cooperation among
Asakuchi-city, NAOJ, and Tokyo Tech.
This research is based on data collected at the Subaru Telescope, which is operated by the National Astronomical Observatory of Japan. 
Some of the data presented herein were obtained at the W. M. Keck Observatory, which is operated as a scientific partnership among the California Institute of Technology, the University of California and the National Aeronautics and Space Administration. The Observatory was made possible by the generous financial support of the W. M. Keck Foundation.
\\
\software{\texttt{astropy}: \citet{Astropy_2013,Astropy_2018},
\texttt{BANYAN-\rm{$\Sigma$}}: \citet{Gagne2018},
\texttt{CHARIS-DRP}: \citet{Brandt2017}, \texttt{CHARIS-DPP}: \url{https://github.com/thaynecurrie/charis-dpp}; \citet{Currie2020b},
\texttt{orvara}: \citet{Brandt_2021_ORVARA}, 
}

%% file: mcmc_results_new.tex
\begin{deluxetable*}{lclc}
\tablewidth{0pt}
\tablecaption{MCMC Orbit Fitting Results \label{tab:mcmc_results}}
\tablehead{
        {\bf Parameter} & 
        {\bf Median and 16--84th Percentile} & 
        {\bf 95\% Credible Interval} &
        {\bf Prior}
}
\startdata
\multicolumn{3}{c}{Fitted parameters} \\ 
\hline
RV Jitter (m s$^{-1}$) & $<$ 0.013\tablenotemark{a} & (0.0, 12.1)  & $1/\sigma_{\rm{Jit}}$ (log-flat)\\
$M_{\rm{pri}}$ ($M_{\rm{\odot}}$) & ${1.30}_{-0.10}^{+0.10}$ & (1.10, 1.49) & $\mathcal{N}(1.3, 0.1)$\\
$M_{\rm{sec}}$ ($M_{\rm{Jup}}$) & ${27.8}_{-5.4}^{+8.4}$ & (19.2, 49.9)  & $1/M_{\rm{sec}}$ (log-flat) \\
a (au) & ${17.5}_{-3.8}^{+7.2}$ & (12.4, 38.0)  & $1/a$ (log-flat)\\
$\sqrt{e}\sin{\omega}$ & ${0.09}_{-0.42}^{+0.39}$ & ($-$0.62, 0.73)  & uniform\\
$\sqrt{e}\cos{\omega}$ & ${0.29}_{-0.85}^{+0.52}$ & ($-$0.90, 0.95)  & uniform\\
Inclination ($\arcdeg$) & ${104.8}_{-6.9}^{+15}$ & (92.2, 155.5)  & $\sin{i}$ ($i$ = 0--180)\\
PA of ascending node $\Omega$ ($\arcdeg$) & ${49.4}_{-8.0}^{+170}$ & (36.8, 228)  & uniform\\
Mean longitude at 2010.0 ($\arcdeg$) & ${188}_{-49}^{+60}$ & (5, 355)  & uniform\\
Parallax (mas)\tablenotemark{b} & ${23.109}_{-0.028}^{+0.028}$ & (23.052, 23.166)  & $\mathcal{N}(\varpi_{\rm{Gaia}}, \sigma_{\varpi,\rm{Gaia}})$ \\
\hline
\multicolumn{3}{c}{Derived parameters}\\
\hline
Orbital period (yrs) & ${63}_{-20}^{+43}$ & (38, 203)  & \\
Argument of periastron $\omega$ ($\arcdeg$) & ${157}_{-132}^{+179}$ & (4, 357)  & \\
Eccentricity $e$ & 0.52 $\pm$ 0.35 & (0.03, 0.97)  & \\
Semi-major axis (mas) & ${404}_{-89}^{+167}$ & (286, 878)  & \\
Periastron time $T_{\rm{0}}$ (JD $-$ 2400000) & ${63360}_{-1419}^{+5307}$ & (58700, 93300)  & \\
Mass ratio &  ${0.0204}_{-0.0040}^{+0.0065}$ & (0.014, 0.037)  & \\
\enddata
\tablenotetext{a}{The maximum likelihood RV jitter is zero.}
\tablenotetext{b}{The uncertainty includes the Gaia uncertainty \citep{Gaia_2021_Summary} in quadrature with the standard deviation in maximum likelihood parallaxes from the chains.}
\tablecomments{$\mathcal{N(\mu, \sigma)}$ represents a Gaussian with mean $\mu$ and variance $\sigma^2$.}
\label{tab:posteriors}
\end{deluxetable*}